\documentclass{kluwer}    

\input epsf.sty

\def\Te{T_{\rm e}}

\begin{document}                              
\begin{article}
\begin{opening}         
\title{Disk--corona interactions in soft spectral states of black hole
          binaries}
\author{Piotr T. \surname{\.{Z}ycki}}  
\runningauthor{Piotr \.{Z}ycki}
\runningtitle{Soft states of black hole binaries}
\institute{N.~Copernicus Astronomical Center, Warsaw, Poland, 
(ptz@camk.edu.pl)}
\date{27 October 2000}

\begin{abstract}
Re-analysis of some archival X--ray data of black hole binaries
in soft spectral states is presented. Results of detailed spectral
fitting, including proper physical description of the X--ray reprocessed
component, indicate that in most cases the soft, thermal disk emission
cannot be described by a simple multi-color blackbody model. 
Additional Comptonization of those seed photons provides an acceptable
description of the soft component, implying a presence of a warm 
($k\Te \sim 1$ keV),
mildly optically  thick ($\tau \sim 10$) phase of accreting plasma.

Combined spectral/temporal analysis of GS~1124-68 during 'flip-flop'
transitions in VHS suggests that the optically thick accretion flow can proceed
in (at least) two modes giving the same time-integrated behaviour 
(i.e.\ energy spectra) but very different timing behaviour.
\end{abstract}

\keywords{black holes, accretion disk, X-rays}

\end{opening}           

\section{The Comptonized soft component}

Accreting black holes, both stellar (including the Galactic $\mu$--quasars)
and supermassive, show strong, soft,
thermal components in their EUV--X-ray spectra, when accreting above certain
fraction of Eddington rate. This component is thought to
come from an optically thick accretion disk, and its spectrum is usually
modeled as a simple, multi-color blackbody. For re-examining a number
of such spectra, I will use the spectral models described in \citeauthor{ZDS98}
\shortcite{ZDS98}.

\subsection{The Intermediate State}

An example of IS is the May 18th observation of GS~1124-68 (Nova Muscae 1991).
Using first the frequently employed but unphysical model, 
{\sc diskbb + smedge(powerlaw)}, gives an unacceptable fit, 
$\chi^2_{\nu} = 38/25$.
Adding a Gaussian line at 6.4 keV does {\em not\/} improve the fit, unless
the line width is $\sim 10$ keV ($\chi^2_{\nu} = 28/23$). Allowing the line
energy to be free gives $\chi^2_{\nu} = 21/23$ for $E = 4\pm 0.5\,$keV.
These values have little to do with Fe spectral features, but rather indicate 
that it is the continuum shape that is not properly modeled.
Using now the {\sc diskbb + thComp + rel-repr} model gives very bad fit,
$\chi^2_{\nu} = 308/24$. Replacing the {\sc diskbb} component with
a Comptonized blackbody ({\sc comptt}) results in a dramatic improvement,
$\chi^2_{\nu} = 13/22$. Thus the best fit is obtained for the proper 
physical 
description of X--ray reprocessing and the {\em Comptonized blackbody\/}
model for the soft component. The same effect is observed in GRO J1655-40, 
GS~2000+25 and XTE J1550-564 (\.{Z}ycki et al., in preparation). 

\subsection{The High State}

\begin{figure*} 
\hfil\parbox{0.49\textwidth}{
\epsfysize = 4. cm
\epsfbox[50 320 550 720]{zycki_TIII-06_fig1a.ps}
}\hfil
\parbox{0.49\textwidth}{
\epsfysize = 4. cm
\epsfbox[50 400 550 750]{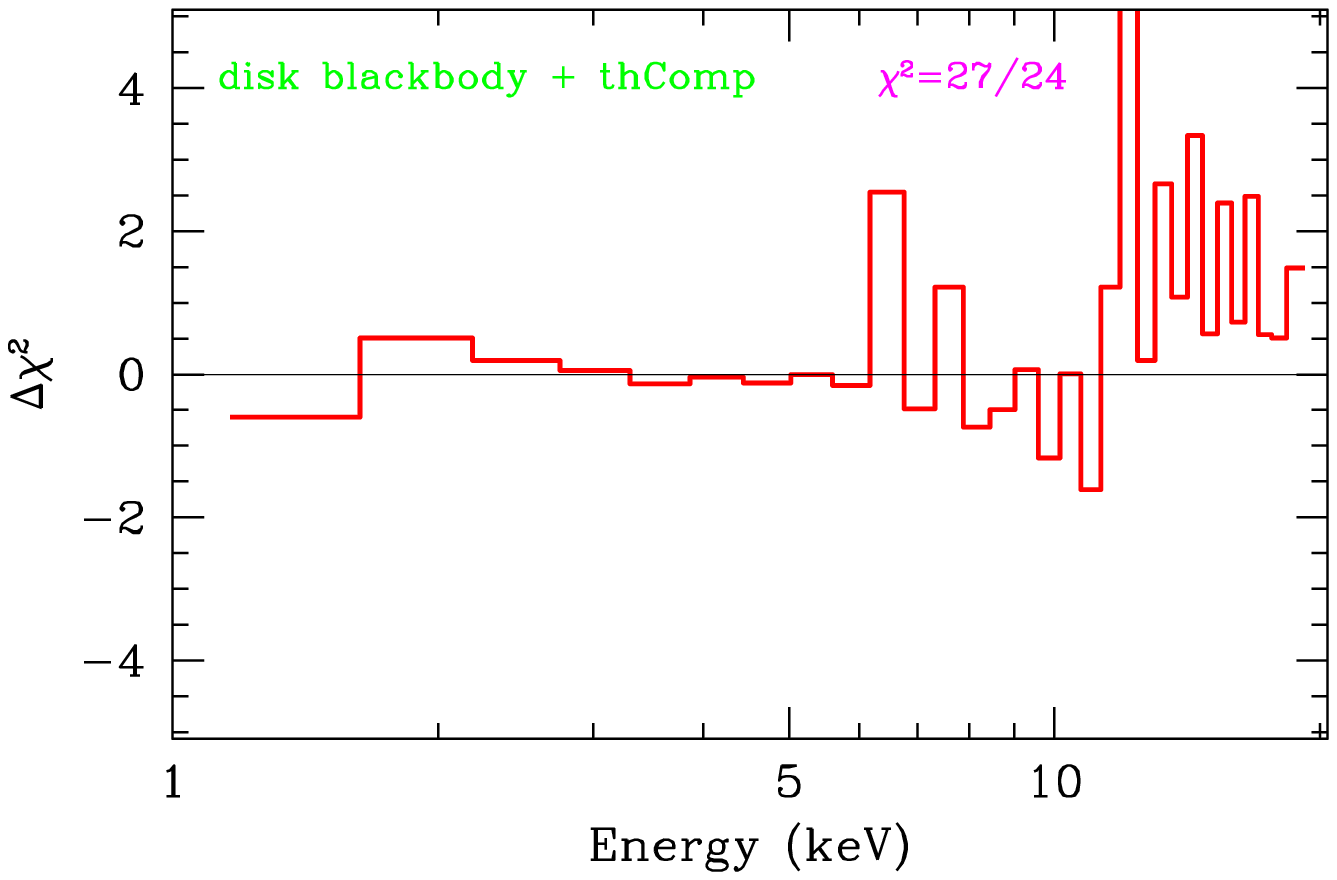}
}\hfil
\caption[]{Spectrum of Nova Muscae 1991 in HS, showing a 
soft thermal component broader than a disk blackbody.}
\label{zycki_TIII-06:fig1}
\end{figure*}

Nova Muscae 1991 was in the High State when observed on Mar 18th.
The spectrum shows a strong soft component, and the hard component
is negligible  \cite{Ebisawa}. Modeling the spectrum with
the {\sc diskbb + thComp } model gives $\chi^2_{\nu} = 27/24$
(see Fig.~\ref{zycki_TIII-06:fig1}). The Comptonized component is
steep and makes a significant contribution to the soft component.
Again, the soft component has to be described by a model {\em broader\/}
than the multi-color disk blackbody.

\subsection{The Very High State}

On Jan 11th, while still on the rise to the peak of the outburst, Nova Muscae
1991, showed a spectrum consisting of two Comptonized components: the softer
one again {\em cannot\/} be described by the {\sc diskbb} model. It can be 
modeled as Comptonization of $\approx 0.3$ keV blackbody by a
warm, $k \Te \approx 3$ keV, optically thick, $\tau\sim 7$, plasma. 

\subsection{Physical origin of the Comptonization}

The additional Comptonization seems to be a common
feature of disk-dominated spectral states of accreting black holes.
There are two possible scenarios to explain it. The first one is 
Comptonization in
a ``hot skin'' on top of the accretion disk. Such a hot layer could not 
be heated by X--ray irradiation (as postulated for the low/hard state),
but it would require enhanced viscous energy dissipation compared to the
bulk of the disk. The second possibility is that the electron energy
distribution in the Comptonizing plasma is a hybrid, thermal--non-thermal one,
 rather than a pure thermal (Maxwellian) distribution. In fact,
the {\sc compPS} model (J. Poutanen, private communication) using
the hybrid distribution gives very good fits to  the HS and IS data sets: 
$\chi^2_{\nu} = 17/24$ for IS, $\chi^2_{\nu} = 16/23$ for HS,
but it fails for VHS.

\section{Inner disk oscillations in GS~1124-68 in VHS?}

Shortly before reaching the peak of outburst Nova Muscae 1991 showed
'flip-flop' transitions \cite{Takizawa}: the flux in the 2--20 keV
band changed by $\approx 20\%$. Spectral and timing analysis during this
period suggests that the source were undergoing re-structuring of the inner
accretion disk.

\subsection{Spectral analysis}

\begin{figure*}[b]
\hfil\parbox{0.5\textwidth}{
\epsfxsize = 0.45\textwidth
\epsfbox[10 320 600 720]{zycki_TIII-06_fig2a.ps}
}\hfil
\parbox{0.45\textwidth}{
\epsfxsize = 0.4\textwidth
\epsfbox[10 240 600 710]{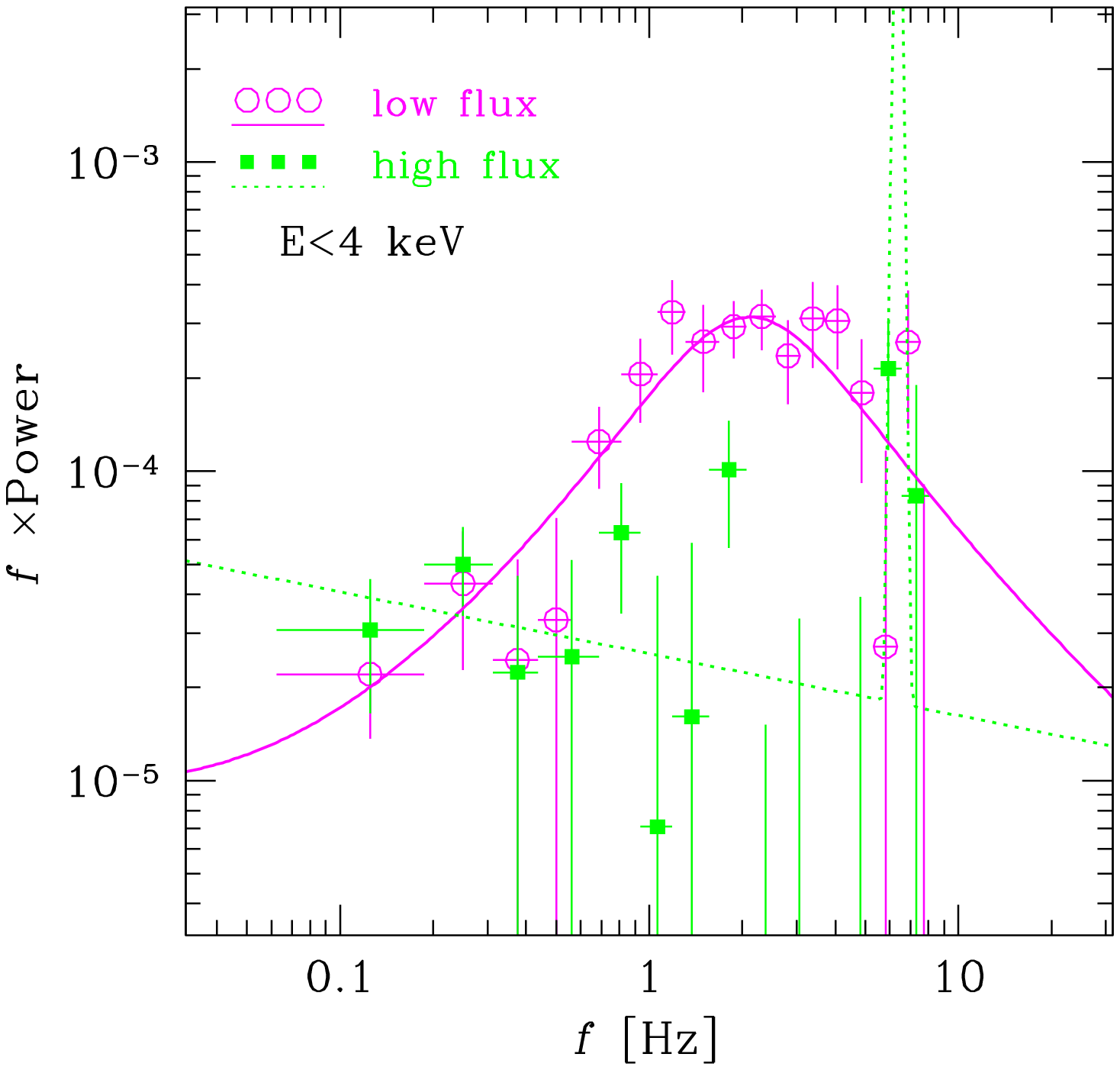}}
\caption[]{Energy and power spectra of Nova Muscae 1991 on Jan 14th (VHS), 
when the source showed 'flip-flop' transitions, 
probably related to re-structuring of the accretion disk. Two spectra
in each panel are for two flux states.}
\label{zycki_TIII-06:fig2}
\end{figure*}

Energy spectra both in the high- and low-flux state can be adequately
described by the {\sc diskbb + thComp + rel-repr} models (no additional 
Comptonization of the soft component is required). The best-fit model
spectra differ most at energies 5--10 keV: the differences are small both
at the lowest and highest energies (Fig.~\ref{zycki_TIII-06:fig2}). 
The difference spectrum can be described
as Comptonization of single blackbody emission, but it cannot be described 
as Comptonized {\em disk\/} blackbody.

\subsection{Timing analysis}

X--ray power density spectra show both energy and flux
dependence. The soft X--rays ($E < 4$ keV) PDS is very different in the
two flux states (Fig.~\ref{zycki_TIII-06:fig2}). In the high flux state
the PDS resembles that of typical HS, i.e.\ a power law $f^{-\alpha}$ with
$\alpha\approx 1.2$, although there is also a narrow QPO at $\approx 8$ Hz. 
However, in the low flux state the PSD has a Lorenzian shape peaking
(in the $f\times Power$ representation) at $\approx 2$ Hz. 

The PDS of the hard X--ray component ($E>8.5$ keV) is similar in both flux
states except for the QPO energy which is higher in the low flux state.

\section{Discussion}

Interactions between accretion disk and corona in the soft spectral states
of accreting black holes are clearly complex. One outcome of those
interactions is a presence of warm, mildly optically thick plasma
modifying the disk thermal emission. Proper modeling of the resulting
spectra is important e.g.\ for  correct determination of system parameters.

The 'flip-flop' transitions observed in VHS of GS~1124-68 close to
the peak of its outburst suggest that the optically thick accretion
flow can proceed in two modes giving the same time-integrated characteristics.
One mode, usually observed in HS, shows  the $f^{-1}$-like PDS, 
i.e.\ a scale-invariant behaviour, 
perhaps realized in magnetic turbulence \cite{Kawaguchi}.
The other mode  clearly possesses certain spatial/temporal 
scale, since the PDS is described by a Lorenzian, i.e.\ it corresponds
to exponential shots of a single time scale.

\acknowledgements
This work  was supported in part by KBN grants 
no.\ 2P03D01816 and 2P03D01718.

\end{article}


\begin{thebibliography}{}


\bibitem[\protect\citeauthoryear{Ebisawa et al.}{1994}]{Ebisawa}
Ebisawa, K. et al.
\newblock {Spectral evolution of the bright X-ray nova GS 1124-68 
(Nova Muscae 1991) observed with GINGA.}
\newblock {\em P.A.S.J.}, 46:375--394, 1994.

\bibitem[\protect\citeauthoryear{Kawaguchi et al.}{2000}]{Kawaguchi}
Kawaguchi T. et al.
\newblock {Temporal $1/f^{\alpha}$ fluctuations from fractal magnetic 
fields in black-hole accretion flow.}
\newblock {\em P.A.S.J.}, 52:1-5, 2000.

\bibitem[\protect\citeauthoryear{Takizawa et al.}{1997}]{Takizawa}
Takizawa, M. et al.
\newblock {Spectral and temporal variability in the X-ray flux of GS 1124-683,
 Nova Muscae 1991.}
\newblock {\em Ap.J.}, 489:272--283, 1997.

\bibitem[\protect\citeauthoryear{\.{Z}ycki et al.}{1998}]{ZDS98}
\.{Z}ycki, P.~T., Done, C. and Smith, D.~A.
\newblock {Evolution of the accretion flow in Nova Muscae 1991.}
\newblock {\em Ap.J.Letter}, 496:L25-L28, 1998.


\end{thebibliography}
\end{document}